%% file: acl_latex.tex
\documentclass[11pt]{article}

\usepackage[preprint]{acl}

\usepackage{times}
\usepackage{latexsym}
\usepackage{booktabs}
\usepackage{xcolor}
\usepackage{tabularx}
\usepackage[T1]{fontenc}

\usepackage[utf8]{inputenc}

\usepackage{microtype}

\usepackage{inconsolata}

\usepackage{graphicx}
\usepackage{times}
\usepackage{latexsym}
\usepackage{amsmath}
\usepackage{enumitem}
\usepackage{amssymb}
\usepackage{tcolorbox}
\usepackage[T1]{fontenc}
\newcommand{\uniticon}[1]{\raisebox{0.6ex}{\includegraphics[height=2ex]{#1}}}
\usepackage[utf8]{inputenc}

\usepackage{microtype}
\usepackage{multirow}
\usepackage{inconsolata}
\usepackage{algorithm}
\usepackage{algpseudocode}
\usepackage{svg}
\usepackage{booktabs}
\title{Privacy-R1: Privacy-Aware Multi-LLM Agent Collaboration via Reinforcement Learning}

\author{
    \textbf{Zheng Hui\uniticon{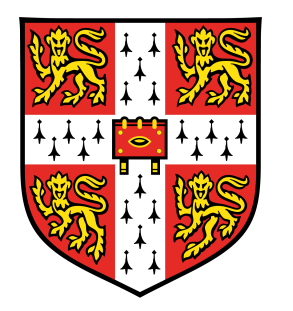}},
    \textbf{Yijiang River Dong\uniticon{image/cam.png}},
    \textbf{Sanhanat Sivapiromrat \uniticon{image/cam.png}}
    \\
    \textbf{Ehsan Shareghi\uniticon{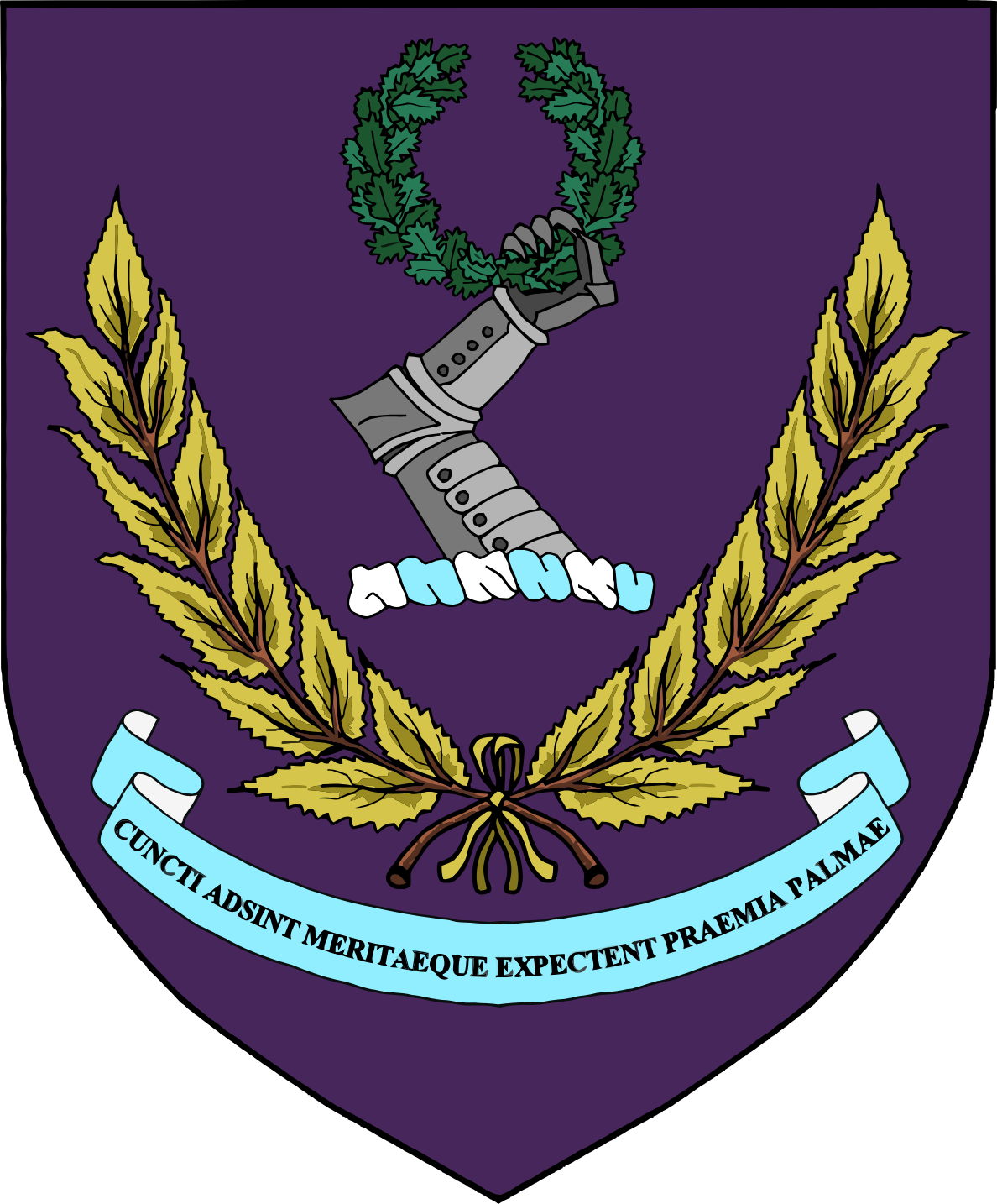}},
    \textbf{ Nigel Collier\uniticon{image/cam.png}}
    \\
    \uniticon{image/cam.png} University of Cambridge,
    \uniticon{image/ucl.png} University College London
    \\
    \{zh2483\}@columbia.edu,
    \\
    \{{yd358, ss3229, es776, nhc30\}}@cam.ac.uk
}

\begin{document}
\maketitle
\input{content/abs}
\input{content/intro}
\input{content/related}
\input{content/problem_formulation}

\input{content/method}
\input{content/exp}

\input{content/result}
\input{content/limitation}
\input{content/ethic}

\bibliography{custom}
\clearpage
\appendix

\input{content/appendix}
\label{sec:appendix}

\end{document}

%% file: content/abs.tex
\begin{abstract}

When users submit queries to Large Language Models (LLMs), their prompts can often contain sensitive data, forcing a difficult choice: Send the query to a powerful proprietary LLM providers to achieving state-of-the-art performance and risk data exposure, or relying on smaller, local models guarantees data privacy but often results in a degradation of task performance. Prior approaches have relied on static pipelines that use LLM rewriting, which shatters linguistic coherence and indiscriminately removes privacy-sensitive information, including task-critical content. We reformulate this challenge (Privacy-Conscious Delegation) as a sequential decision-making problem and introduce a novel reinforcement learning (RL) framework called Privacy-R1 to solve it. Our framework trains an agent to dynamically route text chunks, learning a policy that optimally balances the trade-off between privacy leakage and task performance. It implicitly distinguishes between replaceable Personally Identifiable Information (PII) (which it shields locally) and task-critical PII (which it strategically sends to the remote model for maximal utility). To validate our approach in complex scenarios, we also introduce a new medical dataset with high PII density. Our framework achieves a new state-of-the-art on the privacy-utility frontier, demonstrating the necessity of learned, adaptive policies for deploying LLMs in sensitive environments. Dataset can be found at: https://github.com/zackhuiiiii/Privacy-R1.
\end{abstract}

%% file: content/intro.tex
\section{Introduction}

\begin{figure}[h!]
    \centering
    \includegraphics[width=1\linewidth]{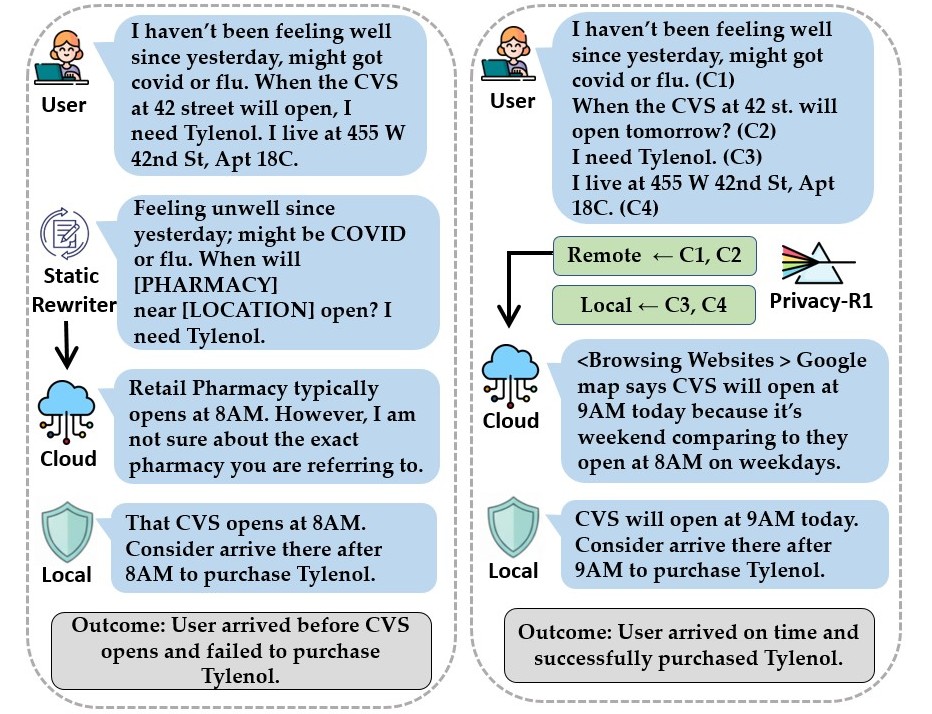} 
    \caption{A static rewriter (left) fails by redacting task-critical information. Our Privacy-R1 framework (right) succeeds by learning a policy to route sensitive and non-sensitive query chunks to the appropriate model, preserving both privacy and task performance.}
    \label{fig:f1}
    \vspace{-0.6cm}
\end{figure}

As users increasingly rely on Large Language Models (LLMs) \cite{brown2020language} across a wide range of domains, their applications now extend far beyond basic text generation. These include synthetic data generation \cite{hui-etal-2024-toxicraft, hui2025toxilabopensourcellmsgenerate}, professional and creative writing \cite{chakrabarty2024creativity}, educational support and tutoring \cite{razafinirina2024pedagogical}, legal drafting \cite{kolt2026legal}, personal assistance systems \cite{li2024personal, dong2026steer} and memory-augmented systems \cite{huang2026rethinking,hui-etal-2026-toward, zhang2026memorycd} that adapt to user preferences over time. As a result, user prompts often contain sensitive information, ranging from personal identifiers and confidential business data to health and financial details \cite{ai-etal-2024-defending}.
 This presents a dilemma: sending a query to a powerful, state-of-the-art remote API risks the exposure private data, while processing it with a secure, locally-hosted smaller model that may yields a lower-quality response. This conflict between utility and privacy is a primary obstacle to deploying LLMs in critical domains like healthcare and finance \cite{rathod2025privacy, hui2025trident}, making the development of trustworthy solutions a crucial area of research.
To formally address this trade-off, recent work has proposed the task of Privacy-Conscious Delegation \cite{li-etal-2025-papillon}. The goal is to create a system where a secure local model acts as an intelligent proxy, leveraging the power of a remote API to fulfill a user's request without leaking their Personally Identifiable Information (PII). Initial approaches to this task have employed a static, monolithic prompt rewriting pipeline. This paradigm attempts to paraphrase the entire user query to remove all PII before sending it to the remote API. However, this strategy is inherently brittle. By treating the entire query as a single unit and removing all PII, it fails to make nuanced decisions, leading to two catastrophic failure modes. The first is discourse coherence failure, where removing entities severs critical linguistic links within the text. The second is utility collapse as illustrated in Figure \ref{fig:f1}, where the PII is integral to the user's goal, and its removal renders the query unanswerable.
The core limitation of these approaches is that they are based on a fixed, context-blind policy. We argue that effective delegation is not a one-shot text manipulation task, but rather a sequential decision-making problem where the optimal action depends on both the immediate context and the overall user objective. To solve this, we introduce a novel reinforcement learning (RL) framework that explicitly learns to balance the competing pressures of privacy and performance. Our framework trains a stateful policy agent to make granular, chunk-by-chunk routing decisions. By optimizing for a reward function that combines downstream task performance with a penalty for privacy leakage, our agent learns a dynamic and pragmatic policy. It learns when to shield information and when a calculated privacy risk is necessary to achieve a high-quality outcome for the user. Our primary contributions are threefold:
\begin{itemize}[leftmargin=1em, itemsep=0.5pt, topsep=2pt]
\item We introduce a novel RL framework call Privacy-R1 featuring a lightweight stateful policy agent that dynamically route content between local and remote models. By optimizing a composite reward, it implicitly learns to differentiate between replaceable PII (which it shields) and task-critical PII (which it may strategically reveal for a significant utility gain).
\item We contribute a new, challenging medical PII dataset, featuring a higher density of PII entities. This benchmark is designed to stress-test the contextual reasoning capabilities of privacy-preserving systems.
\item Our experiments show that Privacy-R1 outperforms baselines, charting a new state-of-the-art on the privacy-utility frontier. The performance gains are especially pronounced on our more complex dataset, underscoring the necessity of adaptive, learned frameworks for building the next generation of trustworthy AI systems.
\end{itemize}

%% file: content/related.tex
\section{Related Works}

\paragraph{Training-Time vs. Inference-Time Privacy}
While much privacy research focuses on training-time risks like data memorization and federated learning \cite{kandpal2022deduplicating, carlini2022quantifying, wang2025llm, ruzzetti-etal-2025-private, dong2026value}, protecting user data at inference time is a critical challenge \cite{mireshghallah2024trust}. Classical approaches \cite{herwanto2021named, mahendran2021privacy, pamarthi2024ai} often rely on Named Entity Recognition (NER) models to perform static redaction method that frequently degrades semantic coherence and downstream task utility. The task of Privacy-Conscious Delegation \cite{li-etal-2025-papillon} was proposed to use LLM to paraphrase the entire user query to remove PII. \textbf{Our work accepts this task definition but argues that a static, global rewrite is suboptimal. We instead propose a dynamic, routing approach.}

\paragraph{Reinforcement Learning for Privacy-Aware Control}

Research at the intersection of RL and privacy has largely concentrated on two main directions. The first focuses train an RL to protect sensitive information against inference or reconstruction attacks \cite{andreoletti2020privacy, qiao2023offline, mo2024security,tan2025federated}. The second line of work applies RL to generative models, where rewards are formulated to balance the quality or utility of generated outputs (e.g., text, images) with privacy metrics~\cite{10483278, saha2025cloud,hore2025deep}. 
In contrast, we are not primarily defending against an attacker or modifying the text itself; instead, we train an agent to learn a \textbf{meta-policy for information routing}. To our knowledge, this is the first work to use RL to learn a stateful, dynamic delegation policy that optimally balances the privacy-utility trade-off at a sub-prompt level.

%% file: content/problem_formulation.tex
\begin{figure*}[h!]
    \centering
    \includegraphics[width=1\linewidth]{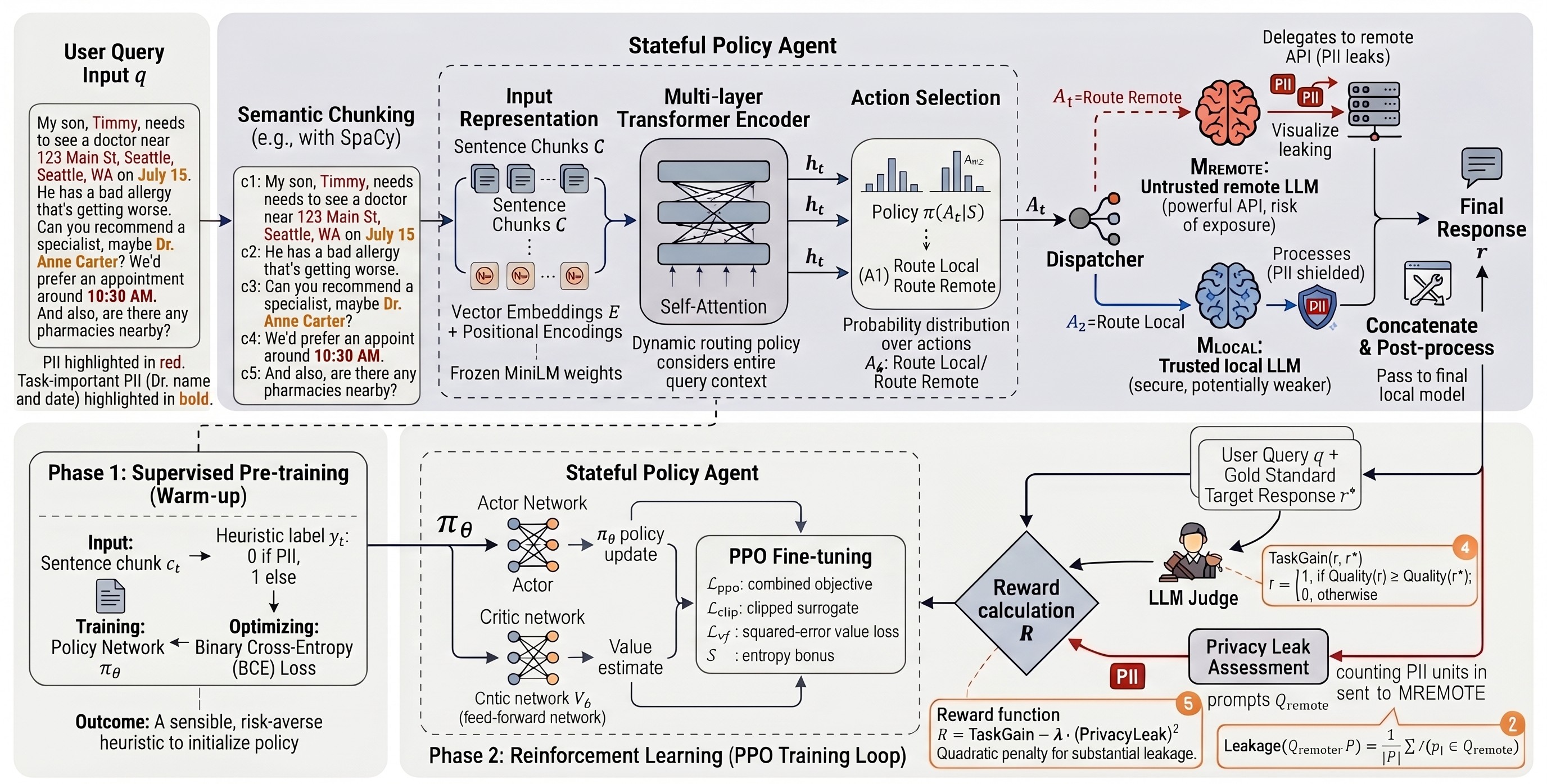} 
    \caption{Privacy-R1 Framework.}
    \label{fig:f2}
    \vspace{-0.6cm}
\end{figure*}

\section{Problem formulation}

\label{sec:problem_formulation}
We address the task of \textbf{Privacy-Conscious Delegation} \cite{li-etal-2025-papillon}. The task is designed to navigate the trade-off between the performance of powerful, remote LLMs and the security of local, trusted models. The setup involves two models:
\vspace{-0.1cm}
\begin{itemize}
    \item $M_{\text{LOCAL}}$: A trusted, secure LLM model with potentially weaker capabilities, which can be run locally or on a private server.
    \vspace{-0.2cm}
    \item $M_{\text{REMOTE}}$: An untrusted, powerful LLM model accessed via Proprietary API.
\end{itemize}
\vspace{-0.1cm}
\noindent The system is given a user query $q$, which may contains a set of one or more Personally Identifiable Information (PII) units, denoted as $P = \{p_1, p_2, \dots, p_k\}$. The fundamental goal is to generate a final response $r$ that is of high quality, while ensuring that the information delegated to $M_{\text{REMOTE}}$ contains as little PII as possible.
\noindent To quantify success in this task, we adopt the evaluation metrics proposed in the original work:

\paragraph{Quality Preservation} The performance of a system is measured by its ability to produce a response $r$ that is comparable in quality to a gold-standard target response, $r^*$. This target is defined as the output from the powerful $M_{\text{REMOTE}}$ when given the original, unaltered query $q$. The quality is determined by a scoring function, $J_{\text{qual}}$:
\begin{equation}
    \text{Quality}(r) = J_{\text{qual}}(r, r^*)
\end{equation}
This function is typically implemented using an LLM-as-a-judge to assess the semantic equivalence and utility of the generated response relative to the target.

\paragraph{Privacy Leakage} The privacy cost is measured as the fraction of PII units from the original query that are exposed to the untrusted remote model. Let $Q'_{\text{remote}}$ be the complete set of all prompts synthesized and sent to $M_{\text{REMOTE}}$ during the delegation process. The leakage is then formally defined as:
\begin{equation}
    \text{Leakage}(Q'_{\text{remote}}, P) = \frac{1}{|P|} \sum_{i=1}^{k} \mathbb{I}(p_i \in Q'_{\text{remote}})
\end{equation}
where $\mathbb{I}(\cdot)$ is the indicator function, which evaluates to 1 if the PII unit $p_i$ is present in any prompt sent to the remote model, and 0 otherwise.

\noindent The central challenge of Privacy-Conscious Delegation is thus to design a system that can effectively navigate the tension between these two competing objectives, maximizing the quality score while minimizing the leakage score.

%% file: content/method.tex
\section{Privacy-R1}
\label{sec:method}

As shown in Figure ~\ref{fig:f2}, to solve the problem defined in Section~\ref{sec:problem_formulation}, we reframe Privacy-Conscious Delegation as a sequential decision-making problem. We introduce a RL framework called \textbf{Privacy-R1} short for \textbf{  \underline{PRIVACY}  preservation policy agent for dynamic delegation via \underline{R}einforcement Learning} that trains a policy agent to learn an optimal, dynamic routing policy. This section details the architecture of our agent, the design of our reward function, and the training procedure used to optimize the policy.

\subsection{Framework Overview}
Our framework processes a user query $q$ in three main stages. First, we perform semantic chunking to segment the query into a sequence of meaningful units. Second, our Stateful Policy Agent processes this sequence to make a routing decision for each chunk. Finally, the chunks are dispatched, and their outputs are combined to form the final response.

\paragraph{Semantic Chunking}
The first step is to segment the user query $q$ into a sequence of coherent units for the policy agent. We employ a semantic chunking strategy. Specifically, we use a sentence segmentation model from SpaCy to split the query $q$ into a sequence of sentences $C = (c_1, \dots, c_n)$. This ensures that each input to our agent is a linguistically complete thought, which provides a much cleaner learning signal and preserves the contextual relationships necessary for learning a robust routing policy.

\paragraph{Dynamic Routing and Response Generation}
The Policy Agent processes the entire sequence of chunks $C$ to assign an action $A_t$ (route local or remote) to each chunk $c_t$. The chunks are then dispatched to the corresponding models ($M_{\text{LOCAL}}$ or $M_{\text{REMOTE}}$), and their individual outputs are concatenated and pass to local model in order to form the final response $r$.
\subsection{The Policy Agent}
To effectively model the complex, long-range dependencies within a user query, our policy agent is implemented using a lightweight \textbf{Transformer} architecture. This design allows the agent to consider the entire query context when making a decision for any individual chunk. The parameters of this Transformer agent are the target of our training procedure.

\paragraph{Input Representation} We use a pre-trained MiniLM model with its weights \textbf{frozen} to serve as a high-quality feature extractor. The full sequence of sentence chunks $C$ is first converted into a sequence of dense vector embeddings $E = (e_1, \dots, e_n)$. To provide positional context, we add standard sinusoidal positional encodings to these embeddings.

\paragraph{Contextualization with Self-Attention} The sequence of embeddings $E$ is then passed through a multi-layer Transformer encoder. The self-attention mechanism computes a new representation, $h_t$, for each chunk $c_t$ that is contextualized by all other chunks in the sequence:
\begin{equation}
    (h_1, \dots, h_n) = \text{Transformer}(e_1, \dots, e_n)
\end{equation}
This global context is critical for making informed decisions, such as understanding that a pronoun in the final chunk refers to a PII entity in the first.

\paragraph{Action Selection} For each contextualized output vector $h_t$, a shared linear layer followed by a softmax activation produces the probability distribution over the two possible actions. This distribution is our policy $\pi$ for each step:
\begin{equation}
    \pi(A_t | S) = \text{Softmax}(W \cdot h_t + b)
\end{equation}
where the state $S$ implicitly represents the entire sequence of chunks. During training, an action $A_t$ for each chunk is sampled from this distribution to encourage exploration.

\subsection{Reward Design: Balancing Privacy and Performance}
\label{sec:reward_design}
The reward signal is calculated after a full sequence of actions has been taken. The reward function $R$ explicitly models the privacy-utility trade-off.

\paragraph{TaskGain}
This term corresponds to the $\text{Quality}(r)$ metric and serves as the primary performance signal for our agent. Following the established protocol from Papillon~\cite{li-etal-2025-papillon}, we treat \texttt{TaskGain} as a \textbf{binary signal}. An LLM-as-a-judge determines if the system's final response $r$ is of 'equivalent or better quality' than the target response $r^*$. The reliability of this specific automated evaluation judge for this task has been previously validated with human studies in Papillon, which found substantial agreement between the LLM judge and human annotators. By adopting this validated judge, we ensure a reliable reward signal for our agent. The reward is formally defined as:
\[
\scalebox{0.8}{$
\text{TaskGain}(r, r^*) =
\begin{cases}
1, & \text{if } \text{Quality}(r) \ge \text{Quality}(r^*) \\
0, & \text{otherwise.}
\end{cases}
$}
\]
To further mitigate known biases, we present the responses to the judge in a randomized order during the reward calculation process.

\paragraph{PrivacyLeak} This term corresponds to the $\text{Leakage}(Q'_{\text{remote}}, P)$ metric, measuring the fraction of PII units exposed to $M_{\text{REMOTE}}$.

\paragraph{Modeling Catastrophic Risk with a Non-Linear Penalty}
To reflect the accelerating nature of real-world privacy risk, we introduce a \textbf{quadratic penalty} for leakage. This penalizes substantial leakage far more harshly than minor, incidental exposure. Our final reward function is:
\begin{equation}
    R = \texttt{TaskGain} - \lambda \cdot (\texttt{PrivacyLeak})^2
\end{equation}
The hyperparameter $\lambda$ controls the agent's aversion to privacy risk. By default, we set $\lambda=5.0$. We also evaluate different levels of $\lambda$ in \ref{sec:lambda_sensitivity}.

\subsection{Policy Training}
We train the policy agent in two phases: a supervised pre-training phase to provide a strong initialization, followed by a reinforcement learning fine-tuning phase to optimize for the reward.

\paragraph{Phase 1: Supervised Pre-training (Warm-up)}
Training an RL agent from a random initialization is challenging due to the sparse and delayed nature of the reward signal. We therefore pre-train our policy network by treating it as a sequence-to-sequence binary classifier, teaching it to mimic a sensible, risk-averse heuristic.
For each sentence chunk $c_t$ in our training corpus, we generate a heuristic label $y_t$: if the chunk contains any PII, $y_t=0$ (route local); otherwise, $y_t=1$ (route remote). The policy network $\pi_{\theta}$ is then trained to predict this sequence of labels. The forward pass computes a probability $p_t = \pi_{\theta}(A_t=1 | S)$ for each chunk. The parameters $\theta$ of the Transformer agent are optimized by minimizing the binary cross-entropy (BCE) loss, summed over all chunks in the sequence. We use the Adam optimizer for this phase. The resulting pre-trained agent, $\pi_{\theta_{\text{sft}}}$, serves as the starting point for RL fine-tuning, already equipped with a conservative policy.

\paragraph{Phase 2: Reinforcement Learning with PPO}
We fine-tune the pre-trained policy using Proximal Policy Optimization (PPO) \cite{schulman2017proximal}. The policy agent serves as the \textbf{actor} ($\pi_{\theta}$), and a separate feed-forward network serves as the \textbf{critic} ($V_{\phi}$).
The training is iterative. In each iteration, we first perform a rollout step, where the actor policy interacts with a batch of queries to collect trajectories of states, actions, and log-probabilities. For each completed trajectory, we compute the final episodic reward $R$ and then calculate the advantage for each step, $\hat{A}_t = R - V_{\phi}(s_t)$, where $V_{\phi}$ is the critic's value estimate. These advantages, which indicate the relative quality of an action, are normalized for training stability. 
The actor and critic are then updated by optimizing a combined objective function. This objective consists of the PPO policy loss, a value function loss for the critic, and an entropy bonus to encourage exploration. The complete objective function to be maximized is:
\begin{equation}
\begin{split}
    \mathcal{L}_{\text{PPO}}(\theta, \phi) = \mathbb{E}_t \big[ & L_t^{\text{CLIP}}(\theta) - c_1 L_t^{\text{VF}}(\phi) \\
    & + c_2 S[\pi_{\theta}](s_t) \big]
\end{split}
\end{equation}
where $L_t^{\text{CLIP}}(\theta)$ is the PPO clipped surrogate objective that constrains the magnitude of the policy update. It is defined as:
\begin{equation}
    L_t^{\text{CLIP}}(\theta) = \min(r_t(\theta)\hat{A}_t, \text{clip}(r_t(\theta), 1-\epsilon, 1+\epsilon)\hat{A}_t)
\end{equation}
Here, $r_t(\theta)$ is the probability ratio between the new and old policies. The other terms are the squared-error value function loss $L_t^{\text{VF}}(\phi)$ and the policy's entropy $S$. This process allows the agent to safely and efficiently learn a sophisticated routing policy that maximizes the expected reward.

%% file: content/exp.tex
\section{Experimental Setup}

\subsection{Datasets}
\paragraph{PUPA}
To benchmark our method against prior work, we first train and evaluate on the \textbf{PUPA} dataset \cite{li-etal-2025-papillon}. PUPA is constructed from real-world user-LLM interactions and contains naturalistic PII across a variety of general-domain topics. The dataset is divided into two primary subsets used for training and evaluation: \textbf{PUPA-New} and \textbf{PUPA-TNB}. We use the official splits and statistics as reported by the authors for our experiments.

\paragraph{Med-PCD}
While PUPA provides a valuable general-domain benchmark, specialized domains like healthcare present unique challenges, including a higher density of interconnected PII and more severe consequences for privacy breaches. To stress-test our framework in such a scenario, we introduce the \textbf{Medical Privacy-Conscious Delegation (Med-PCD)} dataset.
Med-PCD is based on publicly available \textbf{MedDialog} \cite{zeng-etal-2020-meddialog}, which contains anonymized patient-doctor diagnostic conversations. To create Med-PCD, we employed GPT5 to synthetically yet realistically inject a diverse set of PII into these anonymized dialogues. We used a few-shot prompting strategy (see Appendix~\ref{app:prompt} for the full prompt) to guide the model in weaving entities such as patient names, doctor names, clinic locations (e.g., hospitals, departments), specific dates, and medical record numbers (MRNs) into the conversational turns. The goal was not merely to insert PII, but to create coherent and contextually rich narratives where the PII is naturally integrated.

A critical component for our RL framework is the ground-truth target response $r^*$, which is used to calculate the \texttt{TaskGain} reward. To generate these targets, we submitted each complete, PII-injected query from Med-PCD to a powerful proprietary model (GPT5). This response, generated with full access to all PII, represents the quality ceiling that our privacy-preserving system aims to match.
To rigorously assess the quality and realism of our synthetic generation process, we conducted a manual validation study. Three annotators evaluated a randomly selected subset of 240 instances for contextual coherence and logical consistency. The study revealed a high success rate, with an average of 98.8\% of instances being judged as pass. This validation confirms the reliability of our automated data creation pipeline. The inter-annotator agreement was substantial (Fleiss' Kappa $\kappa=0.89$), indicating that our quality criteria were consistently interpreted. The full details of our annotation guidelines and validation study are provided in Appendix~\ref{app:validation}.
 The final Med-PCD dataset consists of 1020 instances, which we split into a training set of 816 instances, and test sets of 204 instances. Table \ref{tab:dataset_stats} provides a statistical comparison between PUPA and Med-PCD, highlighting the increased complexity of our new dataset. Table~\ref{tab:medpcd_examples} illustrates the transformation from an original anonymized dialogue to an injected Med-PCD instance, showcasing the complexity introduced.

\begin{table}[h!]
\centering

\scalebox{0.9}{
\begin{tabular}{lcc}
\toprule
 \textbf{Metrics}& \textbf{PUPA} & \textbf{Med-PCD (Ours)} \\
\midrule
\# Instances & 901 & \textbf{1020} \\
Domain & General & \textbf{Medical} \\
Avg. \# PII & 2.90 & \textbf{4.6} \\
Avg. Query Len. & 1352.0 & \textbf{1533.2} \\
Avg. Resp. Len. & 1553.7 & 1920 \\
\bottomrule
\end{tabular}}
\caption{Statistical comparison of the datasets. The PUPA column shows weighted averages across its splits. Avg length in Char. Med-PCD is designed to be more challenging, with longer queries and a significantly higher density of PII per instance.}
\label{tab:dataset_stats}
\vspace{-0.5cm}
\end{table}

\definecolor{pii_color}{RGB}{211, 47, 47} 
\newcommand{\pii}[1]{\textcolor{pii_color}{\textbf{#1}}} 

\begin{table*}[ht!]
\centering
\small

\begin{tabularx}{\textwidth}{lX}
\toprule
\textbf{Stage} & \textbf{Dialogue Turn (Patient's Message)} \\
\midrule
\textbf{Original (from MedDialog)} & traveled 2wks ago from fl. to pa. 68 wf. has had fever of 100, chills at night and some coughing for 5ds. tested negative for flu and x rays of lungs were clear. coronavirus? scarce \\
\midrule
\textbf{After Injection (Med-PCD)} & My mother, \pii{Carol}, 68, returned from her trip two weak ago. She flew on \pii{UA2401} from \pii{Orlando} to \pii{phil} on \pii{10/21}. For the last 5 days, she's had a fever of 100 and night chills. Her recent flu test at the \pii{Jefferson Health clinic} on \pii{Chestnut St} was negative and x rays of lungs were clear. Could this be coronavirus? \pii{She does not have insurance}. \\
\bottomrule
\end{tabularx}
\caption{An example illustrating the creation of a Med-PCD instance. The original, anonymized text from MedDialog is transformed by synthetically injecting multiple, contextually-aware PII entities (highlighted in red) to create a realistic and challenging scenario.}
\label{tab:medpcd_examples}
\vspace{-0.35cm}
\end{table*}

\subsection{Models}

\paragraph{LLM Configurations}
Following prior work \cite{li-etal-2025-papillon}, we test all methods with a fixed remote model, \texttt{GPT-4o-mini} \cite{achiam2023gpt} as $M_{\text{REMOTE}}$, while systematically varying the local model, $M_{\text{LOCAL}}$. We select a diverse suite of recent open-source models for $M_{\text{LOCAL}}$: \textbf{Llama-3.1-8B-Instruct} \cite{touvron2023llama}, \textbf{Llama-3.2-3B-Instruct} \cite{grattafiori2024llama}, \textbf{Llama-3.2-1B-Instruct} \cite{grattafiori2024llama}, \textbf{Mistral-7B-Instruct-v0.3} \cite{jiang2023mistral7b}, and \textbf{Qwen2-7B-Instruct} \cite{yang2025qwen3}. 
\paragraph{Baselines}
We compare our proposed framework against four strong baselines for each of the local model configurations. \textbf{Always-Local} and \textbf{Always-Remote} establish the performance bounds by exclusively using one model. \textbf{PAPILLON} \cite{li-etal-2025-papillon}, a static, monolithic prompt rewriting pipeline. Finally, to isolate the benefit of reinforcement learning, we include our \textbf{Heuristic Router (SFT-only)}, which is our policy agent after only completing the supervised pre-training phase.

\subsection{Evaluation and Implementation}
\paragraph{Metrics and Protocol}
We evaluate all systems on two primary metrics: \textbf{Quality Preservation (\%)}, the percentage of responses judged equivalent or better than a target, and \textbf{Privacy Leakage (\%)}, the average fraction of PII units exposed to the remote model.

\paragraph{Implementation Details}
Our framework is implemented in PyTorch, using SpaCy for sentence segmentation. Our policy agent is a 2-layer Transformer operating on frozen embeddings from \texttt{all-MiniLM-L6-v2}. The SFT (warm-up) phase is executed for 1 epochs with a batch size of 32 and a learning rate of $3 \times 10^{-4}$. Subsequently, RL with PPO is performed with a learning rate of $1 \times 10^{-5}$ with batch size 64. Training is conducted for a maximum of 256 steps with a clipping parameter $\epsilon=0.2$ and an entropy bonus coefficient of 0.01. All experiments are run on NVIDIA H200 GPUs.

%% file: content/result.tex
\begin{table*}[ht!]
\centering

\scalebox{0.8}{
\begin{tabular}{l ccc ccc}
\toprule
& \multicolumn{3}{c}{\textbf{PUPA-TNB}} & \multicolumn{3}{c}{\textbf{Med-PCD}} \\
\cmidrule(lr){2-4} \cmidrule(lr){5-7}
\textbf{Method / System} & \textbf{Qual. (\%) $\uparrow$} & \textbf{Leak. (\%) $\downarrow$} & $\Delta (\%)$ & \textbf{Qual. (\%) $\uparrow$} & \textbf{Leak. (\%) $\downarrow$} & $\Delta (\%)$ \\
\midrule
GPT-4o-mini [Unredacted] & 88.2 & 100.0 & – & 91.5 & 100.0 & – \\
GPT-4o-mini [Redacted] & 77.2 & 0.0 & – & 72.3 & 0.0 & – \\
\midrule
\multicolumn{7}{l}{\textit{M$_{\text{LOCAL}}$: Llama-3.2-1B-Instruct}} \\
\quad Always-Local & 41.2 & 0.0 &  & 25.5 & 0.0 &  \\
\quad PAPILLON & \underline{58.0} & \underline{39.3} &  & \underline{45.1} & \underline{42.5} &  \\
\quad Heuristic Router (SFT-only) & 52.5 & 2.0 &  & 40.8 & 3.5 &  \\
\quad \textbf{Privacy-R1} & \textbf{62.5} & \textbf{25.0} & \textbf{+4.5 / +14.3} & \textbf{75.3} & \textbf{18.2} & \textbf{+30.2 / +24.3} \\
\cmidrule{1-7}
\multicolumn{7}{l}{\textit{M$_{\text{LOCAL}}$: Llama-3.2-3B-Instruct}} \\
\quad Always-Local & 57.3 & 0.0 &  & 40.1 & 0.0 &  \\
\quad PAPILLON & \underline{60.9} & \underline{24.9} &  & \underline{58.5} & \underline{28.1} &  \\
\quad Heuristic Router (SFT-only) & 59.1 & 1.8 &  & 55.6 & 3.1 &  \\
\quad \textbf{Privacy-R1} & \textbf{65.2} & \textbf{19.5} & \textbf{+4.3 / +5.4} & \textbf{81.0} & \textbf{15.4} & \textbf{+22.5 / +12.7} \\
\cmidrule{1-7}
\multicolumn{7}{l}{\textit{M$_{\text{LOCAL}}$: Llama-3.1-8B-Instruct}} \\
\quad Always-Local & 71.8 & 0.0 &  & 55.0 & 0.0 &  \\
\quad PAPILLON & \underline{85.5} & \underline{7.5} &  & \underline{82.0} & \underline{9.2} &  \\
\quad Heuristic Router (SFT-only) & 78.5 & 1.0 &  & 70.5 & 2.5 &  \\
\quad \textbf{Privacy-R1} & \textbf{86.1} & \textbf{4.5} & \textbf{+0.6 / +3.0} & \textbf{89.5} & \textbf{5.1} & \textbf{+7.5 / +4.1} \\
\cmidrule{1-7}
\multicolumn{7}{l}{\textit{M$_{\text{LOCAL}}$: Mistral-7B-Instruct-v0.3}} \\
\quad Always-Local & 75.7 & 0.0 &  & 58.3 & 0.0 &  \\
\quad PAPILLON & \underline{77.6} & \underline{11.9} &  & \underline{74.5} & \underline{14.0} &  \\
\quad Heuristic Router (SFT-only) & 76.8 & 1.2 &  & 69.1 & 2.8 &  \\
\quad \textbf{Privacy-R1} & \textbf{80.2} & \textbf{8.1} & \textbf{+2.6 / +3.8} & \textbf{87.9} & \textbf{9.5} & \textbf{+13.4 / +4.5} \\
\cmidrule{1-7}
\multicolumn{7}{l}{\textit{M$_{\text{LOCAL}}$: Qwen2-7B-Instruct}} \\
\quad Always-Local & 76.5 & 0.0 &  & 60.1 & 0.0 &  \\
\quad PAPILLON & \underline{78.0} & \underline{15.2} &  & \underline{76.2} & \underline{18.5} &  \\
\quad Heuristic Router (SFT-only) & 77.2 & 1.5 &  & 71.0 & 3.0 &  \\
\quad \textbf{Privacy-R1} & \textbf{81.1} & \textbf{10.5} & \textbf{+3.1 / +4.7} & \textbf{88.4} & \textbf{12.0} & \textbf{+12.2 / +6.5} \\
\bottomrule
\end{tabular}}
\caption{Main results comparing Privacy-R1 against baselines on the PUPA-TNB and Med-PCD test sets across a diverse suite of local models ($M_{\text{LOCAL}}$). Best result in each block is in \textbf{bold} and the second best is \underline{underlined}. $\Delta$ indicates the absolute improvement (in percentage points) of Privacy-R1 over the next best method (PAPILLON). For leakage, a positive $\Delta$ denotes a greater reduction.}
\label{tab:main_results}
\vspace{-0.35cm}
\end{table*}

\section{Results and Analysis}
\label{sec:results}

\begin{figure*}[t]
  \centering
  \includegraphics[width=1\linewidth]{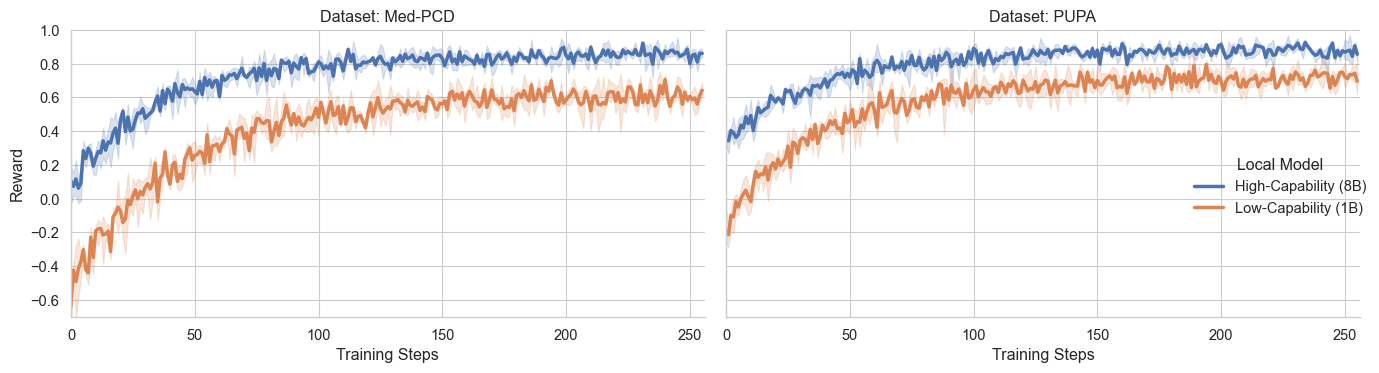}
  \caption{Learning curves showing the reward during PPO training on Med-PCD (left) and PUPA (right).}
  \label{fig:reward_curve}
  \vspace{-0.35cm}
\end{figure*}
In this section, we present the empirical evaluation of our RL framework. We first compare its performance against all baselines on both the PUPA and our new Med-PCD. Subsequently, we provide an analysis of the learning dynamics and conduct a series of ablation studies to dissect the specific contributions of our method's key components.

\subsection{Main Results}
Our primary goal is to demonstrate that a learned, dynamic policy can achieve a better balance between Quality Preservation and Privacy Leakage than static approaches. The main results across a diverse suite of local models are summarized in Table~\ref{tab:main_results}.
The findings are clear: our RL framework consistently establishes a new state-of-the-art on the privacy-utility frontier. On the general-domain \textbf{PUPA benchmark}, our method achieves a better Pareto-optimal point across all local model configurations. For instance, with Llama-3.1-8B, it improves Quality Preservation by +0.6\% while reducing Privacy Leakage by -3.0\% over PAPILLON. This highlights the benefit of dynamic routing even in standard scenarios.

Our Privacy-R1 achieves a better privacy and utility balance in more PII dense \textbf{Med-PCD dataset}. Here, the performance gap widens dramatically. The quality of the baselines, particularly PAPILLON and Always-Local, drops significantly due to the data's complexity. In contrast, our RL framework's quality remains very high. With Llama-3.1-8B, our framework outperforms PAPILLON by a remarkable +7.5\% in Quality Preservation while still leaking -4.1\% less PII. This is because the high density of interconnected PII in Med-PCD is particularly detrimental to static rewriting baselines, which often remove task-critical context. Our stateful agent, in contrast, learns to navigate this complexity.
Furthermore, the results show that the benefit of our routing agent is most pronounced when the local model is weakest. With the 1B local model on Med-PCD, the Always-Local baseline quality collapses to 25.5\%. In this setting, our RL framework provides a massive +30.2\% quality improvement over PAPILLON, confirming that when the cost of routing locally is high, the intelligence of the routing policy becomes the single most important factor in the system's success.

\subsection{Analysis of Learning Dynamics}
To provide insight into the learning process, Figure \ref{fig:reward_curve} plots the reward per training steps. We compare the learning dynamics of agents paired with a high-capability local model (Llama-3.1-8B) and a low-capability one (Llama-3.2-1B) on both datasets.

The learning curves reveal two key insights. First, all agents demonstrate stable learning, with the episodic reward consistently increasing before converging, confirming the effectiveness of our PPO training setup. Second, on both datasets, the agent paired with the high-capability 8B model converges to a significantly higher final reward. This is because the penalty for a ``safe" local action is much lower with a stronger local model, making the overall credit assignment problem easier and the achievable reward higher. The consistently lower reward curves on the Med-PCD dataset, compared to PUPA, visually confirm the increased difficulty of our new benchmark.

\subsection{Ablation Studies}

To isolate the contributions of our key design choices, we conduct a series of ablation studies on the Med-PCD dataset. We use \textbf{Qwen2-7B-Instruct} as the local model to demonstrate generalizability. We compare our full framework (trained with $\lambda=5.0$) against variants where one component is removed or simplified.

\paragraph{Importance of Stateful Context}
To prove the value of our stateful Transformer architecture, we created a \textbf{Stateless Router (MLP)} baseline, where an MLP makes an independent routing decision for each chunk with no memory of surrounding chunks. As shown in Table~\ref{tab:ablation_state}, the stateless agent's performance drops significantly, losing over 13\% in quality compared to our full stateful model. The stateless agent frequently fails on queries with anaphora or cross-chunk dependencies, providing direct evidence that stateful context is critical for this task, regardless of the underlying LLM.

\begin{table}[h!]
\centering

\scalebox{0.9}{
\begin{tabular}{lcc}
\toprule
\textbf{Method} & \textbf{Quality (\%) $\uparrow$} & \textbf{Leakage (\%) $\downarrow$} \\
\midrule
Stateless Router & 75.2 & 11.5 \\
\textbf{Stateful Router} & \textbf{88.4} & \textbf{12.0} \\
\bottomrule
\end{tabular}}
\caption{Ablation on the importance of a stateful agent, using Qwen2-7B as the local model.}
\label{tab:ablation_state}
\vspace{-0.35cm}
\end{table}

\paragraph{Effect of Non-Linear Reward}
Finally, we analyze the impact of our quadratic leakage penalty by training a variant with a standard \textbf{Linear Reward} ($R = \text{TaskGain} - \lambda \cdot \text{PrivacyLeak}$). While the linear reward agent achieves a comparable average performance point (Table~\ref{tab:ablation_reward}), its behavior is far more reckless. We measure this by reporting the percentage of test instances that result in a catastrophic leak (defined as >80\% of PII leaked). The agent trained with our quadratic penalty almost never exhibits this failure mode, confirming that the non-linear penalty is crucial for training a safer, more reliable policy.

\begin{table}[h!]
\centering

\scalebox{0.75}{
\begin{tabular}{lcc}
\toprule
\textbf{Reward} & \textbf{Quality (\%) $\uparrow$} & \textbf{Catastrophic Leaks (\%) $\downarrow$} \\
\midrule
Linear Penalty & 88.1 & 16.2 \\
\textbf{Ours} & \textbf{88.4} & \textbf{1.1} \\
\bottomrule
\end{tabular}}
\caption{Ablation on the effect of the reward function's penalty term, using Qwen2-7B. The quadratic penalty drastically reduces the risk of catastrophic leaks.}

\label{tab:ablation_reward}
\vspace{-0.35cm}

\end{table}

\subsection{Analysis of the Privacy-Utility Trade-off}
\label{sec:lambda_sensitivity}
A key feature of our Privacy-R1 framework is the ability to control the agent's behavior via the privacy-utility trade-off parameter, $\lambda$. This hyperparameter is not meant to be tuned for a single best value; rather, it provides a direct mechanism to generate policies with different levels of risk aversion. To demonstrate the effectiveness of this control, we trained separate Privacy-R1 agents on the Med-PCD dataset with varying values of $\lambda$, using \textbf{Qwen2-7B-Instruct} as the local model.
The results, presented in Table~\ref{tab:lambda_sensitivity}, show a clear and predictable relationship between $\lambda$ and the agent's final policy. As we increase the penalty for privacy leakage (i.e., increase $\lambda$), the agent learns a progressively more conservative policy. The Privacy Leakage drops monotonically from 15.5\% at a low $\lambda$ of 1.0 to a near-zero 1.2\% at a high $\lambda$ of 20.0. This comes at the expected cost of Quality Preservation, which also decreases as the agent is forced to rely more heavily on the weaker local model.
This analysis confirms that $\lambda$ serves as an effective and intuitive control knob. 

\begin{table}[h!]
\centering

\scalebox{0.9}{
\begin{tabular}{ccc}
\toprule
\textbf{$\lambda$} & \textbf{Qual. (\%) $\uparrow$} & \textbf{Leak. (\%) $\downarrow$} \\
\midrule
1.0 & 90.1 & 15.5 \\
2.0 & 89.6 & 13.8 \\
\textbf{5.0 (Default)} & \textbf{88.4} & \textbf{12.0} \\
10.0 & 84.7 & 5.3 \\
20.0 & 79.2 & 1.2 \\
\bottomrule
\end{tabular}}
\caption{Sensitivity analysis of the trade-off parameter $\lambda$ on the Med-PCD dataset, using Qwen2-7B-Instruct
as the local model.}
\label{tab:lambda_sensitivity}
\vspace{-0.35cm}
\end{table}

\section{Conclusion}
\label{sec:conclusion}

In this work, we addressed the challenge of the privacy-utility tradeoff in the task of Privacy-Conscious Delegation. We reframed the task as a sequential decision-making problem and introduced \textbf{Privacy-R1}, a novel reinforcement learning framework. By optimizing a reward function that balances task success with a non-linear penalty for privacy leakage, our agent learns a pragmatic policy that distinguishes between replaceable and task-critical PII. It achieves a better balance of quality and privacy on both PUPA and Med-PCD. This work marks a clear step away towards a future of dynamic, learned policies for building AI systems that are both powerful and trustworthy.

%% file: content/limitation.tex
\section{Limitations}
\label{sec:limitations}

The scope of our work is shaped by several key design choices, which in turn delineate the boundaries of our claims. First, our investigation—consistent with the foundational task definition—is conducted in a single-turn setting. The agent's state and learned policy are self-contained within a single user query and do not persist across multi-turn dialogues. 
Second, the action space of our delegation agent is restricted to a binary decision between a single local model and a single remote model. We do not explore the more complex scenario of delegating among a broader set of specialized local or remote models, each potentially exhibiting different cost, latency, or capability profiles. 
Nevertheless, we believe that extending this framework to support \textbf{multi-turn dialogue privacy preservation} and \textbf{dynamic delegation across multiple specialized models} represents a promising direction for future work.

%% file: content/ethic.tex
\section{ Ethical Considerations}
\label{sec:ethics}

The primary motivation for this research is to enhance user privacy in the age of powerful, API-based Large Language Models. Our goal is to develop a framework that mitigates the risk of exposing sensitive user data to third-party services, thereby enabling the safer use of state-of-the-art AI technologies in critical domains. We have considered the ethical implications of our work throughout the research process, from dataset creation to the potential deployment of our method.

\paragraph{Dataset Creation}
Our new benchmark, Med-PCD, was created with privacy as a foremost concern. We built upon the MedDialog dataset, which is a publicly available resource already anonymized to remove real patient information. The Personally Identifiable Information (PII) in Med-PCD is entirely synthetic, generated by a large language model, and does not correspond to any real individuals. The resulting dataset will be handled responsibly and made available to researchers in a manner that respects its intended use for studying privacy-preserving systems.

\paragraph{Intended Use and Risks}
Privacy-R1 is designed as a risk mitigation framework, not a formal privacy guarantee. It is crucial to understand that our system reduces privacy leakage but does not eliminate it entirely. The reinforcement learning agent is trained to make optimal trade-offs based on its reward function and may still choose to leak PII if it perceives a sufficient gain in task performance. Therefore, Privacy-R1 should not be deployed in applications where a 100\% guarantee against data exposure is required.

The tunable hyperparameter, $\lambda$, is a core feature of our responsible design. It provides system developers with a direct and intuitive control knob to set the desired level of risk aversion, allowing them to tailor the policy to be more or less conservative based on the specific sensitivity of the application's domain.

\paragraph{Broader Impact}
We believe the broader impact of this research is positive. By demonstrating a more effective method for balancing privacy and utility, our work can encourage the development of hybrid AI systems that are more respectful of user data. This could facilitate the adoption of powerful language models in fields like healthcare and finance, where privacy concerns have traditionally been a major barrier. We encourage the community to build upon this work to develop even more robust and formally verifiable privacy-preserving frameworks.

%% file: content/appendix.tex
\appendix
\section{Prompt for PII Injection}
\label{app:prompt}

Below is the few-shot prompt used to guide the GPT-4o model in generating the Med-PCD dataset. The prompt provides clear instructions, constraints, and a comprehensive list of PII types to ensure the model produces high-quality, coherent, and realistic synthetic data.

\begin{quotation}
\noindent\rule{\linewidth}{0.4pt}
\textbf{System Prompt}

You are an expert data annotator specializing in creating realistic, synthetic medical records for research. Your task is to take an anonymized medical dialogue turn and enrich it by injecting a diverse and creative set of Personally Identifiable Information (PII).

**Instructions:**
1.  Read the anonymized input text carefully to understand the context.
2.  Inject a variety of PII types from the comprehensive list below. Be creative and make the details specific.
3.  Ensure the dialogue remains fluent, natural, and contextually coherent. The PII should feel like a natural part of the conversation or note.
4.  **Crucially, do not change the original medical facts, symptoms, diagnoses, or timelines.** You are only adding personal, logistical, and administrative details.
5.  The injected PII should be logically consistent within the text.

**PII Types to Inject (be creative and specific):**
*   **Personal Identifiers:**
    *   Full Names (e.g., Carol prequel)
    *   Dates of Birth (DOB) (e.g., 05/12/1956)
    *   Medical Record Numbers (MRN) (e.g., JH-48920-C)
    *   Phone Numbers (e.g., 215-555-0182)
    *   Email Addresses (e.g., carol.p@emailservice.com)
*   **Healthcare Specific:**
    *   Doctor/Clinician Names (e.g., Dr. Anya Sharma)
    *   Hospital/Clinic Names (e.g., Jefferson Health, Penn Medicine)
    *   Specific Departments (e.g., Urgent Care, Radiology Department)
    *   Insurance Information (Provider and Policy Number, e.g., Aetna Blue PPO \#789-456-123)
    *   Pharmacy Names (e.g., CVS on Walnut St)
*   **Temporal and Location Information:**
    *   Specific Dates (e.g., October 21st)
    *   Appointment Times (e.g., 3:45 PM)
    *   Full Street Addresses (e.g., 1101 Chestnut St)
    *   Cities / States (e.g., Philadelphia, PA)
*   **Logistical / Contextual Details:**
    *   Travel Information (Flight Numbers, Airlines, e.g., United Flight UA2401)
    *   Workplace Names (e.g., her employer, Comcast Center)
    *   Vehicle Information (e.g., License Plate, if relevant to an accident)

---

\textbf{Example 1:}

\texttt{Anonymized Input:}
\texttt{Patient is a 45-year-old male with a history of hypertension. He came in today for a routine check-up. Blood pressure is slightly elevated. Will adjust his medication and schedule a follow-up in 3 months.}

\texttt{Injected Output:}
\texttt{Patient \textbf{David Chen} (MRN: \textbf{PMH-38192}, DOB: \textbf{03/14/1979}) is a 45-year-old male with a history of hypertension. He came in today, \textbf{November 5th, 2024}, for a routine check-up with \textbf{Dr. Anya Sharma}. Blood pressure is slightly elevated. Will adjust his Lisinopril prescription (sent to the \textbf{Walgreens on El Camino}) and schedule a follow-up at the \textbf{Palo Alto Medical Foundation} in 3 months.}

---

\textbf{Now, complete the following:}

\texttt{Injected Output:}
\noindent\rule{\linewidth}{0.4pt}
\end{quotation}

\section{Human Validation Study of the Med-PCD Generation Process}
\label{app:validation}

To ensure that our automated PII injection pipeline produces high-quality and realistic data, we conducted a rigorous human validation study. This appendix details the methodology and findings of that study.

\paragraph{Study Design}
Three graduate students, proficient in English and familiar with NLP annotation tasks, served as our annotator. They were presented with a randomly selected sample of 240 instances from the generated Med-PCD dataset. For each instance, they were asked to independently provide a "pass" or "fail" judgment based on two core quality criteria.

\paragraph{Annotation Guidelines}
Our guidelines instructed raters to disregard minor grammatical errors, abbreviations, or informalities that were likely present in the original MedDialog text. The focus was strictly on the quality of the PII injection. An instance was judged as a "pass" only if it met both of the following criteria:

\begin{enumerate}
    \item \textbf{Contextual Coherence:} The injected PII fits naturally and logically within the medical context of the dialogue.
    \item \textbf{Logical Consistency:} The injected PII does not create any internal contradictions within the text.
\end{enumerate}

To ensure consistent judgments, we provided clear examples of what constitutes a "fail" for each criterion:

\begin{itemize}
    \item \textbf{Coherence Failure:} The injected PII is contextually inappropriate, irrelevant, or nonsensical.
    \textit{Example:} A doctor's note discussing a patient's blood pressure results suddenly includes an unrelated flight number like "United Flight UA2401".

    \item \textbf{Consistency Failure:} The injected PII contains internal contradictions.
    \textit{Example:} The text mentions a patient's DOB is "\textit{05/12/1956}" but also refers to them as a "25-year-old male".
\end{itemize}

\paragraph{Results and Inter-Annotator Agreement}
The results of the study strongly validated our data generation process. Averaged across the three raters, \textbf{98.8\% of the evaluated instances were judged as a "pass"}, indicating that our pipeline consistently produces high-quality, coherent, and logically sound data.

To measure the reliability of the rating process itself, we calculated the Inter-Annotator Agreement (IAA) using Fleiss' Kappa ($\kappa$), the standard metric for assessing agreement among a fixed number of raters. We achieved a substantial agreement with a calculated $\kappa = 0.89$. This high level of agreement confirms that our quality criteria were clear and that the high success rate is a reliable measure of our dataset's quality.